**Characterization of Mega-Constellation Links for LEO Missions With Applications to EO and ISS Use Cases**

G. Maiolini Capez, M. A. Caceres, C. P. Bridges, S. Frey, R. Armellin, R. Garello, P. Bargellini

---

---

---

# Characterisation of Mega-Constellation Links for LEO Missions with Applications to EO and ISS Use Cases

**G. MAIOLINI CAPEZ[1], M. A. CACERES[2], C. P. BRIDGES[3], S. FREY[2], R. ARMELLIN[4], R. GARELLO[1], (SENIOR MEMBER, IEEE), P. BARGELLINI[5]**

[1]Department of Electronics and Telecommunications, Politecnico di Torino, Turin, Italy
[2]Vyoma GmbH, Darmstadt, Germany
[3]Surrey Space Centre, University of Surrey, Guildford, United Kingdom
[4]Department of Mechanical Engineering, University of Auckland, Auckland, New Zealand
[5]European Space Research and Technology Centre, European Space Agency, Noordwijk, Netherlands

Corresponding author: R. Garello (e-mail: roberto.garello@polito.it).

This study was funded by the European Space Agency, in the frame of the Open Space Innovation Platform implemented through its Discovery & Preparation programme, under contract number 3-16610/20/NL/GLC.

**ABSTRACT** Satellite missions demand ever greater connectivity, especially in the LEO regime. In this paper, we introduce the new mega-constellation services in space paradigm: we show that mega-constellations, deployed to offer innovative services to Earth's users, can provide excellent connectivity to LEO spacecrafts, too. First, we characterise the communication link between space users and the actual OneWeb and Starlink constellations. A full set of results in terms of availability, access duration, Doppler, and path losses as a function of user orbital parameters, identifying optimal user orbits, is provided. The results achieved by a multi-system user able to communicate with both fleets are also presented. The potential improvements available if geostationary constellations are used to complement LEO mega-constellations in a multi-orbit system are discussed, too. Finally, we focus on two LEO use cases, the International Space Station and an Earth Observation Sun Synchronous satellite. All the results demonstrate the numerous advantages of the mega-constellation connectivity solution, which is able to transform LEO spacecrafts into highly responsive nodes of a space-to-space network.

## I. INTRODUCTION

Currently, there is a lot of interest in the satellite industry to transform Low-Earth Orbit (LEO) satellites into 24/7 connected nodes of an Internet-like space network where they can autonomously communicate among themselves, streaming data as it is generated on-board, and thus enabling real-time applications [1], [2]. Proposed methods include enhancing ground station network capabilities while increasing their size [3], moving towards higher frequency bands (i.e., Ka/V/Q bands) and optical telescopes [4].

Moreover, we can also observe two very important trends in the satellite industry. Communication satellite operators have been working towards providing seamless, high throughput, low latency connectivity to terrestrial users worldwide through LEO mega-constellations. At the same time, a lot of research is ongoing aiming at integrating the various LEO, Mid-Earth Orbit (MEO), and Geostationary Orbit (GEO) platforms into an unified multi-orbit system, capable of finely balancing capacity, coverage, cost, throughput, and latency according to customer needs [5]–[11].

In recent years, data relay constellations (typically in GEO, but some in LEO, too) [12], [13] have been envisioned to address such a need through radiofrequency (RF) and optical interfaces, being proposed by academia [14] and industry [15]–[17] alike. Furthermore, recent developments in optical terminal capabilities promise a great increase in capacity and efficiency so far unachievable by most space users, including remote sensing and Earth observation satellites.

Even though extremely interesting for future applications, only few optical data relay services are currently active. Moreover, they may be affected by limited scalability - only a few users can be served by each terminal, and there are only so many terminals a satellite can carry [18]. On the other hand, mega-constellations' RF links are not only already operational but can potentially offer much greater scalability (tens of thousands of concurrent users) and reduced operational complexity. As a consequence, we opt to exploit existing and, above all, cost-effective RF solutions that can be readily put into play.

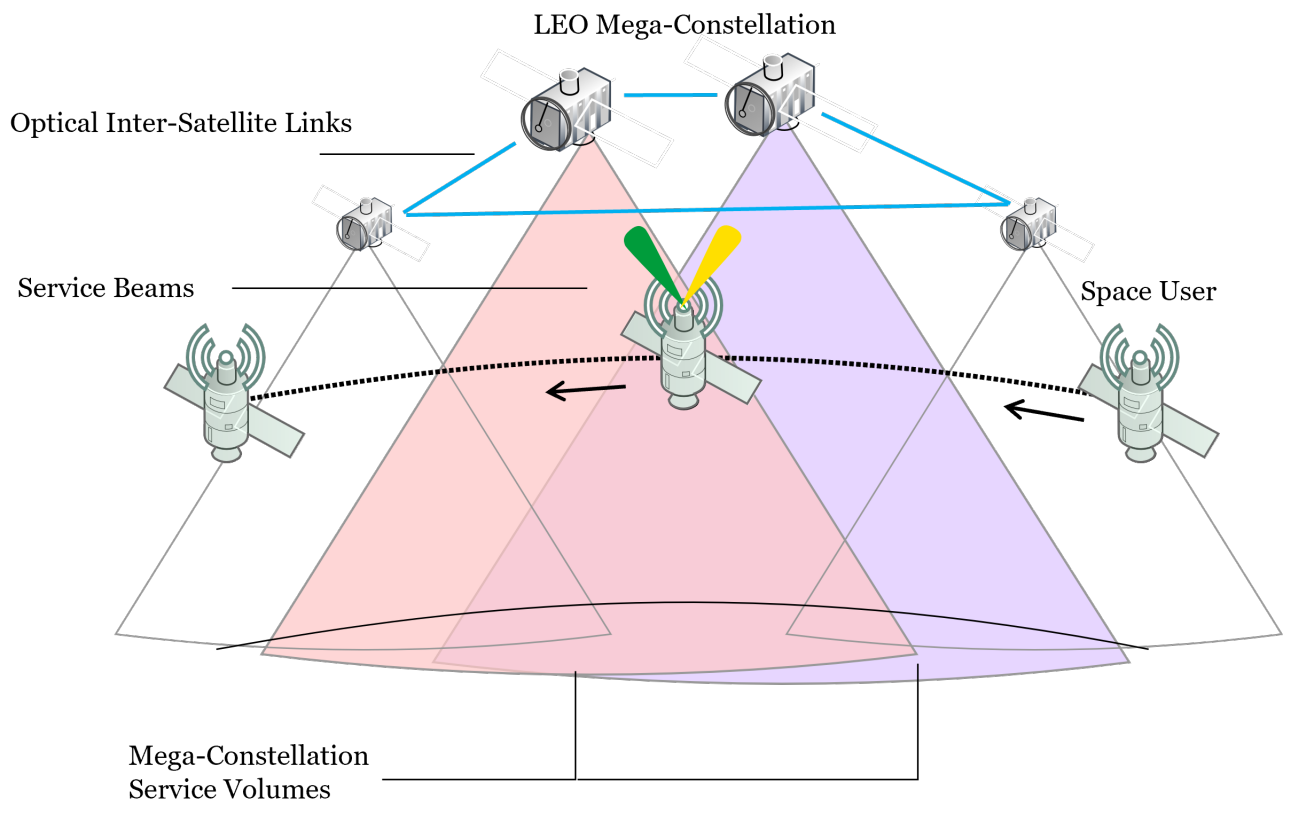

FIGURE 1: Mega-Constellation Services in Space Concept

Thus, we present the new **mega-constellation services in space paradigm**, developed with the support of the European Space Agency (ESA) [19]. The key idea is that mega-constellations, originally designed to offer innovative services to Earth users, can be effectively exploited to connect LEO users, too (Figure 1). (In this paper we denote any user satellite, spacecraft, or object that operates in LEO as a "LEO user", or simply as a "user".)

Building on this paradigm, we provide a full characterization of the communication link between mega-constellations, such as OneWeb and Starlink, and LEO users in terms of coverage probability, path loss, Doppler, pass duration, and access duration. Then, we present the advantages that can be achieved by a multi-system user that can communicate with both fleets. Since in the near future operators will offer integrated GEO and LEO solutions [20], we also provide some insight on the benefits of such a solution to LEO users, by considering the potential improvement in OneWeb's coverage offered by a set of Eutelsat wide-beam Ku-band satellites - selected to reflect an extremely likely integrated solution to be offered as a result from the Eutelsat/OneWeb merger [20].

To the best of the authors' knowledge, a detailed characterisation of mega-constellation links as a function of space user orbital distributions has not yet been conducted in literature prior to this work, much less considering an integrated GEO and LEO solution as the one resulting from the Eutelsat/OneWeb merger. This is a key gap in literature that had yet to be addressed, of relevance even to 6G systems because space users can be anything, including gNodeBs. In closing this gap, we also answer the questions hereto unanswered: "What is space user service volume of mega-constellations? What is the quality of service that a space user can expect? How do the space user orbital parameters impact the quality of service?"

| Acronym | Definition |
|---|---|
| EDRS | European Data Relay System |
| EO | Earth Observation |
| ESA | European Space Agency |
| FCC | Federal Communications Commission |
| FSPL | Free-Space Path Loss |
| GEO | Geostationary Orbit |
| GNSS | Global Navigation Satellite System |
| ISS | International Space Station |
| LEO | Low-Earth Orbit |
| MEO | Mid-Earth Orbit |
| NASA | National Aeronautics and Space Administration |
| RF | Radiofrequency |
| SSO | Sun-Synchronous Orbit |
| TDRSS | Tracking and Data Relay Satellite System |

TABLE 1: List of Acronyms

Yet, the core mega-constellation services in space paradigm presented is much more ambitious than closing a literature gap, it is a novelty; it proposes a new way to look at mega-constellation services similarly to what happened in the navigation world when Global Navigation Satellite System (GNSS) services were first explored for space users. Position, Navigation, and Timing services were initially only available for terrestrial users, but as research progressed by adapting existing receivers and designing new ones, the service volume expanded from Earth to LEO to GEO and beyond [21]–[24].

All the results show that the analyzed solutions can provide a very effective connection for LEO users below 500-km (>80%) or at critical inclinations (i.e., polar) up to 800-km with average path losses ranging from 144 to 175 dB for LEO mega-constellations. To further support our results, we focus on two important use cases: the International Space Station (ISS) and a Sun-Synchronous Orbit (SSO) Earth Observation (EO) satellite in LEO. A complete analysis in terms of coverage, pass duration, visible satellites, Doppler, path loss, elevation angle is provided. All the presented results show that LEO mega-constellations can effectively provide excellent connectivity to a variety of missions, transforming them into low latency nodes of a space network.

The paper is organized as follows. Acronyms are listed in Table 1. Section II introduces the mega-constellations considered in this study as providers of space-based connectivity services. Section III presents the methodology employed in Section V, which characterises space-based connectivity services as a function of space user orbital parameters and service providers. Next, Sections VI and VII explore two representative use cases for such connectivity: the International Space Station and an Earth Observation satellite in Sun-Synchronous Orbit. Finally, Section VIII summarises key results and identifies promising lines of research for future works.

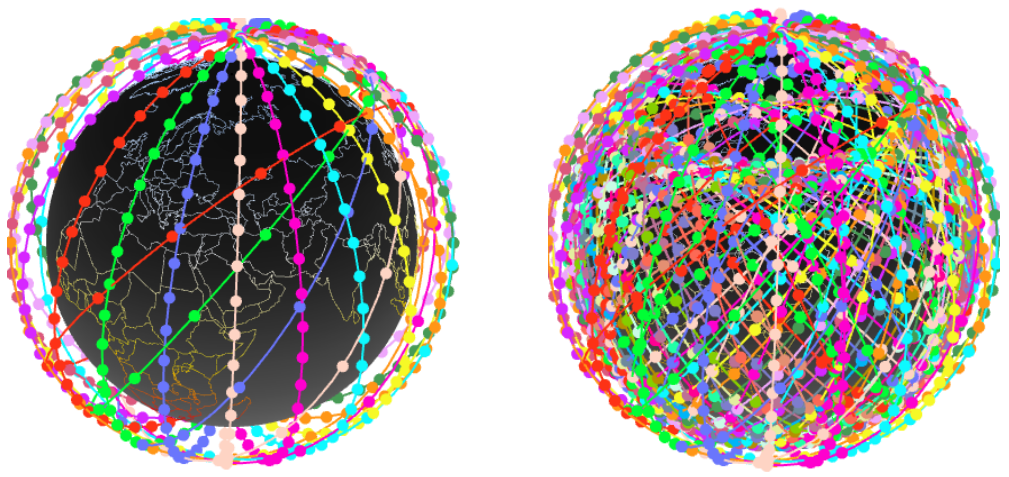

FIGURE 2: OneWeb Phase 1 (left), Starlink Phase 1 (right) LEO Mega-Constellations

## II. SYSTEM DESCRIPTION: MEGA-CONSTELLATIONS AND LEO USERS

As realistic service providers, we consider the OneWeb [25], [26] and Starlink [27], [28] Phase 1 LEO mega-constellations, as per Federal Communications Commission (FCC) filings, and Eutelsat GEO platforms. OneWeb and Starlink aim at providing low-latency broadband services to terrestrial users worldwide using hundreds of satellites with highly directive antennas operating mainly in Ku-band (selected as per FCC filings of deployed user Earth terminals and constellation satellites). Currently, both are in advanced deployment stage, with Starlink being estimated to have nearly 500,000 subscribers [29]. Because these systems have similar characteristics and operate in the same frequency bands, it may be so that in the future, a single terminal may be able to select which network it wants to connect to - just as a cellphone would select a mobile operator. Therefore, we also consider a combined OneWeb + Starlink constellation that would allow users to freely select which operator to use at a given time.

### *A. NETWORKS*

In this phase, OneWeb consists of 716 satellites at 1200 km with 588 satellites across 12 orbital planes at 87.9 degrees and 128 satellites distributed over eight evenly spaced planes at 55 degrees. Starlink instead has 4408 satellites across five orbital shells between 540 and 570 km, each containing several planes, as described in Table 2

| Mega-Constellation | Altitude [km] | Inclination [deg] | Planes | Satellites/Plane | Total |
|---|---|---|---|---|---|
| OneWeb | 1200 | 87.9 | 12 | 49 | 588 |
| OneWeb | 1200 | 55.0 | 8 | 16 | 128 |
| Starlink | 540 | 53.2 | 72 | 22 | 1584 |
| Starlink | 550 | 53 | 72 | 22 | 1584 |
| Starlink | 560 | 97.6 | 6 | 58 | 348 |
| Starlink | 560 | 97.6 | 4 | 43 | 172 |
| Starlink | 570 | 70.0 | 36 | 20 | 720 |

TABLE 2: Mega-Constellation Configuration

Additionally, the 23 Eutelsat GEO satellites considered in the paper to explore the integrated GEO+LEO solution are indicated in Table 3.

### *B. LEO USERS*

As LEO users, we consider a general case and two specific use cases:

| | | | | | |
|---|---|---|---|---|---|
| EUTELSAT 133 WEST A | | | | | |
| 1 26719U 01011A | 22207.65483163 | .00000076 | 00000+0 | 00000+0 0 | 9998 |
| 2 26719 | 3.8069 | 82.4286 0002221 | 45.9245 278.8235 | 1.00273016 78289 | |
| EUTELSAT 5 WEST A | | | | | |
| 1 27460U 02035A | 22207.86438281 -.00000043 | 00000+0 | 00000+0 0 | 9998 | |
| 2 27460 | 2.4102 | 88.9715 0004299 | 32.7417 128.9573 | 1.00272201 73365 | |
| EUTELSAT 7A | | | | | |
| 1 28187U 04008A | 22207.65352362 | .00000105 | 00000+0 | 00000+0 0 | 9997 |
| 2 28187 | 2.2051 | 87.9326 0003884 | 40.7676 271.6781 | 1.00274957 67095 | |
| EUTELSAT 33E | | | | | |
| 1 33750U 09008B | 22207.85119043 | .00000158 | 00000+0 | 00000+0 0 | 9991 |
| 2 33750 | 0.0497 | 10.7166 0003887 112.4049 160.8764 | 1.00271852 49199 | | |
| EUTELSAT 10A | | | | | |
| 1 34710U 09016A | 22207.86924347 | .00000070 | 00000+0 | 00000+0 0 | 9992 |
| 2 34710 | 0.0459 | 6.6953 0005781 118.9906 141.7486 | 1.00271098 27007 | | |
| EUTELSAT 36B | | | | | |
| 1 36101U 09065A | 22207.85119043 | .00000158 | 00000+0 | 00000+0 0 | 9997 |
| 2 36101 | 0.0646 | 11.2899 0004716 113.5616 161.9570 | 1.00271581 46464 | | |
| EUTELSAT 7 WEST A | | | | | |
| 1 37816U 11051A | 22207.82549138 -.00000058 | 00000+0 | 00000+0 0 | 9998 | |
| 2 37816 | 0.0592 | 13.1474 0004385 131.9836 | 89.2170 | 1.00271985 39646 | |
| EUTELSAT 16A | | | | | |
| 1 37836U 11057A | 22207.86924347 | .00000107 | 00000+0 | 00000+0 0 | 9993 |
| 2 37836 | 0.0666 357.4394 0005222 125.9230 150.0960 | 1.00274231 31718 | | | |
| EUTELSAT 21B | | | | | |
| 1 38992U 12062B | 22207.68561337 | .00000133 | 00000+0 | 00000+0 0 | 9999 |
| 2 38992 | 0.0672 357.1924 0002906 123.2259 | 92.3544 | 1.00274719 35589 | | |
| EUTELSAT 70B | | | | | |
| 1 39020U 12069A | 22207.78938233 -.00000042 | 00000+0 | 00000+0 0 | 9993 | |
| 2 39020 | 0.0681 339.6003 0006693 200.2602 119.2627 | 1.00274575 35255 | | | |
| EUTELSAT 117 WEST A | | | | | |
| 1 39122U 13012A | 22207.65797084 -.00000010 | 00000+0 | 00000+0 0 | 9993 | |
| 2 39122 | 0.0268 300.7913 0002662 188.0817 295.5041 | 1.00271998 34186 | | | |
| EUTELSAT 7B | | | | | |
| 1 39163U 13022A | 22207.75443559 | .00000050 | 00000+0 | 00000+0 0 | 9998 |
| 2 39163 | 0.0687 350.9649 0002782 | 52.6043 179.4191 | 1.00272144 33717 | | |
| EUTELSAT 3B | | | | | |
| 1 39773U 14030A | 22207.85882875 | .00000023 | 00000+0 | 00000+0 0 | 9994 |
| 2 39773 | 0.0618 358.4806 0004406 157.8849 100.4260 | 1.00274608 29817 | | | |
| EUTELSAT 115 WEST B | | | | | |
| 1 40425U 15010B | 22207.44311365 -.00000022 | 00000+0 | 00000+0 0 | 9995 | |
| 2 40425 | 0.0034 202.8355 0000308 240.9397 264.9331 | 1.00270767 27146 | | | |
| EUTELSAT 8 WEST B | | | | | |
| 1 40875U 15039B | 22207.86438281 -.00000063 | 00000+0 | 00000+0 0 | 9993 | |
| 2 40875 | 0.0642 358.0344 0005117 | 99.4690 150.1943 | 1.00271942 25394 | | |
| EUTELSAT 65 WEST A | | | | | |
| 1 41382U 16014A | 22207.45305449 -.00000274 | 00000+0 | 00000+0 0 | 9999 | |
| 2 41382 | 0.0476 | 7.7320 0001861 118.5569 275.7374 | 1.00272483 23284 | | |
| EUTELSAT 117 WEST B | | | | | |
| 1 41589U 16038B | 22207.62688292 -.00000009 | 00000+0 | 00000+0 0 | 9999 | |
| 2 41589 | 0.0072 199.0041 0000469 229.2265 344.7213 | 1.00268989 22415 | | | |
| EUTELSAT 172B | | | | | |
| 1 42741U 17029B | 22207.79563142 -.00000004 | 00000+0 | 00000+0 0 | 9990 | |
| 2 42741 | 0.0434 168.3596 0000398 163.4249 | 71.0895 | 1.00269630 19465 | | |
| EUTELSAT 7C | | | | | |
| 1 44334U 19034B | 22207.03695182 | .00000051 | 00000+0 | 00000+0 0 | 9996 |
| 2 44334 | 0.0698 239.4766 0003095 232.5351 212.0227 | 1.00273922 11844 | | | |
| EUTELSAT QUANTUM | | | | | |
| 1 49056U 21069B | 22207.80605059 | .00000128 | 00000+0 | 00000+0 0 | 9997 |
| 2 49056 | 0.0150 354.9819 0001685 129.5126 157.8231 | 1.00271995 | 3674 | | |
| EUTELSAT 36CEXPRESS-AMU1 | | | | | |
| 1 41191U 15082A | 22207.85119043 | .00000158 | 00000+0 | 00000+0 0 | 9998 |
| 2 41191 | 0.0184 | 54.9345 0002548 | 79.1727 152.9083 | 1.00270469 24017 | |
| EUTELSAT 53AEXPRESS-AM6 | | | | | |
| 1 40277U 14064A | 22207.91993869 | .00000100 | 00000+0 | 00000+0 0 | 9991 |
| 2 40277 | 0.0301 235.3385 0000800 325.7639 127.6451 | 1.00273196 28418 | | | |
| EUTELSAT 174A | | | | | |
| 1 28924U 05052A | 22208.14649097 | .00000011 | 00000+0 | 00000+0 0 | 9991 |
| 2 28924 | 0.6057 | 92.4634 0006001 | 33.2606 | 45.7942 | 1.00271787 60730 |

TABLE 3: Eutelsat Network Configuration (Two-Line Elements)

- LEO spacecrafts in near-circular orbits at altitudes below 1200 km and inclination from 0 to 180 deg (inclinations greater than 90-degrees indicate a retrograde orbit).
- An EO satellite in Sun-Synchronous Orbit at an altitude of 500-km.
- The ISS, at a 420-km altitude, 51.6 degree inclination [42].

The two use cases were selected because they represent the most likely user orbits for a variety of applications (i.e., Earth Observation, Remote Sensing, etc.) [41] and benefit from excellent coverage by mega-constellations because of the low user altitude.

## III. METHODOLOGY

### *A. SIMULATION*

Starting from FCC filings of Earth Stations (terminals, gateways) and Space Stations (satellites) of each selected satellite operator [26], [28], that is, official records produced by operators and submitted to the government of the United States for authorization to operate satellite services, we created realistic scenarios around potential space user orbits and terminals that are compatible with operational satellite constellations. By using information provided by the satellite operators themselves, we ensure that the data is robust and faithful to the deployed constellations and their coverage volumes.

To characterise mega-constellation links, an analysis technique had to be selected. Analytical techniques where cover-age is computed considering exact (location-based) or randomly distributed (stochastic) user and satellite positions [30]–[33], are extremely useful to derive initial metrics at reduced complexity since they eliminate the need for precise orbit propagation. However, such a level of abstraction makes it impossible to distinguish and compare the performance of mega-constellations with similar orbital configurations, but different technologies and capabilities. In these instances, numerical analysis is the only solution.

Such a solution has been employed throughout literature to analyse mega-constellation performance [34], [35] and relies on the step-wise propagation of the LEO user and constellation satellite positions, trading-off computational complexity, which scales with the number of LEO users, for numerical accuracy [36]. Thus, we have implemented our own numerical simulator in Python that starting from a set of initial state vectors of positions and velocities, user and constellation satellite positions are iteratively propagated using a SGP4 model [37] over a time window of interest. From their positions, the simulator computes the relative position, range, Doppler offset, and visibility angles between user and constellation satellites, and evaluates whether the visibility angles satisfy constellation and user constraints (as per FCC filings). That is, whether the constellation satellite is in line-of-sight of the user at an elevation equal to or greater than the minimum elevation angle of the user's Zenith-pointing antenna, like a ground station (and vice-versa) [38]. Whenever this condition is satisfied, a possible access is available.

At this point, the simulator applies user policies (i.e., which satellite to select (random or closest), Doppler offset/rate constraints, ...) and computes the relevant (i.e., minimum, mean, maximum) metrics regarding Free-Space Path Loss (FSPL), Doppler rate, coverage probability, number of visible satellites, access duration, among others. The resulting data is binned according to user orbital parameters and plotted for visualization. There is no additional processing performed. To assist readers in reproducing the results, the results of a Monte Carlo run of the simulator for the OneWeb, Starlink, SES, and O3b mPOWER constellations can be found at [39].

Considering that the mega-constellation service volumes encompass hundreds to thousands of kilometers above Earth, that signal characteristics depend on the relative dynamics of the platforms hosting the radio payloads, and taking into account that computational complexity scales with the number of satellites, step size, and simulation horizon, we employ a Monte Carlo [40] strategy with 1000 LEO users uniformly distributed across the main payload planes, and to improve accuracy near constellation shells, where geometric constraints are strongest, we add 100 users at an altitude below and within 100 km of each constellation's maximum altitude ([470 570] km, [1100 1200] km). For the general case of LEO spacecrafts in near-circular orbits we have chosen a random selection among the mega-constellation visible satellites. This allows to focus on the statistical characterization of the channel, without selecting a specific user policy. For the two ISS and EO use cases, where the orbital parameters are fixed, the closest constellation satellite is instead selected, to exploit a more favorable path loss and to ensure the periodicity of the connectivity. The impact of the different selection policies, outside the scope of this paper, is left for future studies (see Section VIII).

Since OneWeb and Starlink are designed to serve terrestrial users within their orbital shell, we only examine LEO missions in near-circular orbits at altitudes below 1200 km as potential users. Also, taking into account that active payloads are overwhelmingly located in LEO at 500-to-550 km altitudes in near-circular orbits with 45-to-60-degree and 80-to-100-degree inclinations [41], as said before we focus on two particular use cases that well represent the most likely user orbits used by many important LEO applications: a SSO EO satellite at an altitude of 500-km and ISS, at a 420-km altitude, 51.6-degree inclination [42]. Moreover, GEO systems such as the European Data Relay System (EDRS) are typically used to relay data from the ISS and ESA's Earth Observation missions. Thus, for comparison purposes, the analysis also discusses the potential services offered by considering one of the world's largest GEO service providers: Eutelsat (a choice that also allows to explore the results of a possible merger Eutelsat/OneWeb).

These simulation scenarios start at 2021-03-20T09:37:29.000Z and run for 24 hours using a 10s-time step for greater resolution. We chose this date because it is an equinox, key for the eventual assessment of Sun-outages, that is, when the Sun is within a satellite's antenna beam. We consider a 25-degree minimum elevation angle (a trade-off between coverage probability, path losses, and antenna constraints; lower angles would result in higher losses and more complex antenna designs, but minimal increase in coverage, while a much higher angle would compromise coverage and access duration) , realistic for both passive (fixed beam) and active antennas (steerable, switchable, or shapeable beams) and, to illustrate some periodic effects due to orbital dynamics, whenever there are at least than two constellation satellites available, the closest constellation satellite is selected for communications to exploit a more favorable path loss. Different strategies (i.e., random selection) could have been employed, but that would come at the cost of losing insight into the periodicity of the connectivity between users and constellation satellites.

### B. DEFINITION OF METRICS

In addition to traditional metrics such as Free Space Path Loss (FPSL), Doppler offset and rate, we define the following metrics:

1) Coverage Probability: Probability that a user can see/be seen by at least one constellation satellite.
2) Pass Duration: How long a pass between a user and a single constellation satellite lasts.
3) (Network) Access Duration: How long at least one constellation satellite is visible by the user, assuming that

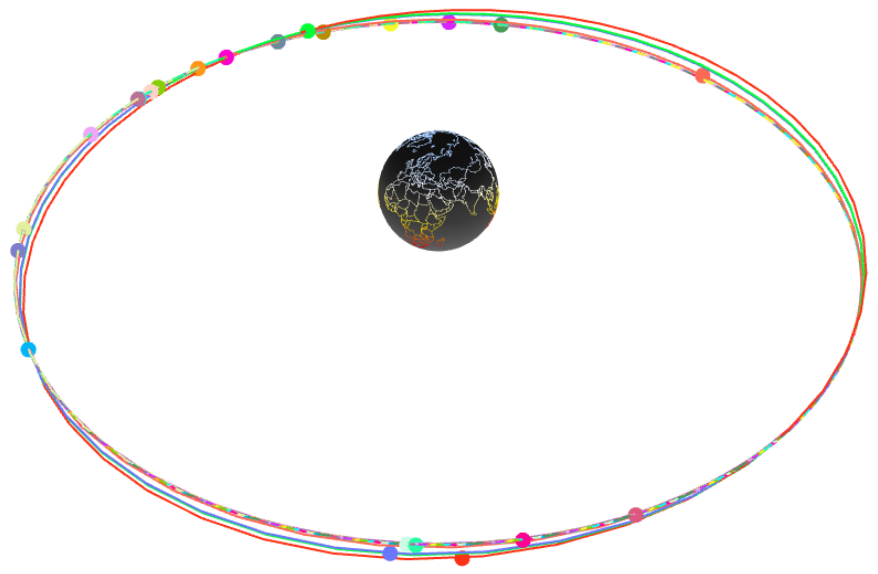

FIGURE 3: Eutelsat Ku-band GEO Constellation (23 satellites)

the user can switch from one satellite to the other with only a brief disconnection (shorter than the simulation time step).

## IV. POTENTIAL ADVANTAGE OFFERED BY A MULTI-ORBIT SYSTEM

In the near future, operators will offer multi-orbit solutions [20]. For this reason, it is useful to investigate the potential gain achievable by integrating our LEO mega-constellations paradigm with GEO satellites. As a case study we have considered Eutelsat's 23 Ku-band wide-beam GEO operational satellites (Figure 3). These satellites offer variety of services, such as data, mobility, and broadband with near-global terrestrial coverage in the Northern and Southern hemispheres, except for polar caps [43].

To study the impact of these GEO satellites on our multi-constellation paradigm, we have used an approximated but effective approach. We focus on the overall service area under the hypothesis that satellites do not irradiate beyond the edge of the visible Earth to minimise interference and power consumption, instead of faithfully representing specific beams, service areas, and capacities, which, unlike LEO mega-constellations, greatly vary between satellites. In this case, the field-of-view of each satellite can be approximated as a 10.5-degree half-cone, which is comparable to other GEO systems such as National Aeronautics and Space Administration (NASA)'s Tracking and Data Relay Satellite System (TDRSS) [44] (13-degree). This first approximation is sufficient to provide a useful analysis of coverage, link, and capacity bounds because it reflects system-level characteristics.

Clearly, the detailed analysis of the GEO satellites impact, the proper selection of their number, band and access techniques, or the study of multi-beam platforms are outside the scope of this paper, which is focused on the characterization of the LEO mega-constellations links, and is left for future studies (see Section VIII.) Anyway, as shown in the next sections, the LEO+GEO results obtained with our model look very interesting and already provide remarkable information.

## V. CHANNEL CHARACTERISATION

In this section, we characterise the main channel parameters of space-based connectivity services as a function of space users' orbital elements and service providers, summarised in Table 4, across all possible LEO user orbits.

| Constellation | OneWeb | Starlink | OneWeb + Starlink | Eutelsat | OneWeb + Eutelsat |
|---|---|---|---|---|---|
| Min. FSPL [dB] | 127.01 | 117.92 | 117.92 | 203.82 | 127.01 |
| Avg. FSPL [dB] | 165.44 | 154.68 | 165.24 | 205.15 | 187.41 |
| Max. FSPL [dB] | 176.73 | 166.85 | 176.73 | 206.32 | 206.32 |
| Max. Doppler Offset [kHz] | 471.79 | 543.36 | 528.38 | 314.28 | 456.40 |
| Avg. Access Duration [min] | 39.32 | 2.77 | 60.96 | 277.82 | 562.64 |
| Max. Access Duration [min] | 1441.00 | 278.00 | 1441.00 | 1441.00 | 1441.00 |
| Overall Coverage [%] | 45.88 | 37.33 | 49.00 | 84.98 | 94.26 |

TABLE 4: Key Constellation Coverage Metrics (25-degree Min. El. Angle, Random Satellite Selection)

### A. AVAILABILITY

Considering a 25-degree minimum elevation angle and selecting the serving satellite at random from those in visibility, Figure 4 shows that as user altitude increases, coverage quickly diminishes and becomes negligible within 50 km of the highest altitude satellites because of the geometric constraints imposed by the user and satellite beamwidths, which determine coverage volume.

OneWeb's coverage starts at 100% at 350 km, falls to 40-to-50% for users at 800-to-850 km, and depends on inclination, with near-polar users (user satellites with 90-to-100 degrees of inclination) having the greatest coverage and equatorial users (user satellites with 0 degrees of inclination) the lowest. Since 82% of OneWeb's satellites are in near-polar orbits and 18% in mid-inclination orbits, equatorial uses frequently cross constellation orbital planes. When the user altitude is below 500 km, its distance to the constellation satellites is sufficiently high to ensure coverage between orbital planes. However, as altitude of the LEO user increases, coverage is degraded because the range between the LEO user and the constellation satellite decrease, narrowing the beams.

Starlink has six times more satellites than OneWeb, at altitudes equal to or below 570 km, making coverage almost independent of user inclination at altitudes below 500 km. Nevertheless, because 72% of Starlink satellites are placed at mid-inclinations, mid-inclination users are always better covered. Unfortunately, most users at 500-to-570 km are too close to constellation satellites and have minimal coverage due to visibility constraints. When both constellations are used, users below 500 km have full coverage regardless their inclination. Above this threshold, coverage is almost exclusively provided by OneWeb.

Overall, Starlink's coverage probability is 37.33% for users at 350-to-550 km altitudes while OneWeb's is 45.88% for users at 350-to-1200 km. Combining both systems increases Starlink coverage by 31.26%, while Starlink only boosts OneWeb's coverage by 6.8% because OneWeb already covers most of Starlink's service volumes due to its higher altitude. Thus, adding Starlink only eliminates some of OneWeb's dead zones due to the angular spacing of its polar orbital planes.

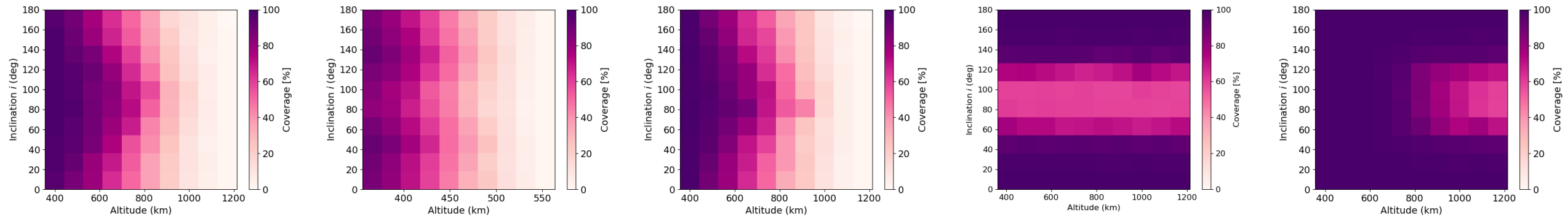

(a) OneWeb (b) Starlink (c) OneWeb+Starlink (d) Eutelsat (e) Eutelsat+OneWeb

FIGURE 4: Satellite Constellation Availability (inclinations greater than 90-degrees indicate a retrograde orbit).

Because of its equatorial inclination, Eutelsat can provide continuous service to low-to-mid inclination users across all altitudes, but cannot offer more than 40-to-50% availability to polar satellites, which experience excellent coverage by OneWeb at altitudes of up to 800 km. Thus, by unifying OneWeb and Eutelsat's GEO platforms into a single integrated constellation with a 94.26% availability, it is possible to provide continuous access to all but polar users at altitudes above 800 km - of great value to most missions.

### B. OPTIMAL USER ORBITAL ELEMENTS

For each constellation, one can identify the optimal user orbital elements. For maximising coverage, users should be below 450 km at 45-to-135-degree inclinations and below 375 km at 40-to-60 degree inclinations for OneWeb and Starlink, respectively. Instead, for at least 50% of coverage, users should be below 800 km at 60-to-120 degree inclinations and 450 km at 30-to-150 degree inclinations for OneWeb and Starlink, respectively. Eutelsat's coverage is independent of user altitude; nearly all users have at least 50% of coverage, with those in equatorial inclinations experiencing full coverage.

### C. ACCESS DURATION

Focusing on the average *daily* access duration, OneWeb can provide 600-to-800 minutes of continuous access for near-polar users below 500 km, while Starlink can only provide up to 16-minutes for users at mid-inclinations below 400 km. These extremely long accesses are possible at low altitudes, where coverage is very high, or whenever there is “coupling” between the user and constellation shells, as for polar users, which move in parallel to the OneWeb's polar satellites.

Figure 5 makes this coupling effect clearer by scaling the access time axis to have a maximum contact duration of 30 minutes for LEO and 120 minutes for LEO/GEO. The ideal inclination to maximise average access duration is close to 90 degrees for OneWeb and 53/127 degrees for Starlink. The farther the inclination is away from these values, the shorter the accesses, which also shrink as altitude rises because of fewer satellites in visibility.

As one can expect, the maximum access time distribution (not shown) reflects the average access distribution; polar OneWeb users at altitudes below 500 km have continuous accesses lasting over 12 hours, while Starlink user accesses usually last fewer than 3 hours. When both constellations are combined, users below 450 km experience longer accesses than with a single constellation due to increased coverage. Similarly, Eutelsat + OneWeb's average access times are several hours long for most users, but are shorter than 20 minutes for high altitude polar users close to OneWeb's shells.

### D. DOPPLER

Figure 6 shows an analytical approximation [45] of the Doppler shift and Doppler rate experienced by a space user at various altitudes when communicating with a constellation satellite at a 1200 km altitude, that is, from the OneWeb constellation, as a function of the user elevation angle. Coverage of space users at those altitudes by Starlink satellites is negligible because of Earth-facing beams. This approximation is used to reduce simulation complexity and improve the Doppler rate estimate at the zenith only, where constellation satellites are rarely present and then only for short windows of time.

While accurate at high elevation angles and useful to illustrate the space user Doppler shift and rates, at lower elevation angles the approximated Doppler shift appears to decrease and deviates from simulation results, as space users also have the well-known "S-curve" Doppler profile usually associated with ground users. Nevertheless, that does not impact the maximum Doppler offset results presented in Table 4 and also discussed below because all results outside from zenith come from simulations.

When the user is at least 10-degrees away from the zenith, the Doppler rate is below 1 kHz/s, but at the zenith, it can reach several kHz/s. While this could potentially be an issue for communication, these are rare events that can be avoided by selecting a different visible satellite to communicate with (if available). Doppler offsets can be as high as 550 kHz, 470 kHz, and 314 kHz for Starlink, OneWeb, and Eutelsat, respectively, as one could expect from orbital dynamics: the higher the satellite altitude, the lower the range rate. Also, the maximum Doppler offset is minimum for equatorial inclinations and increases as the inclination approaches 90 degrees.

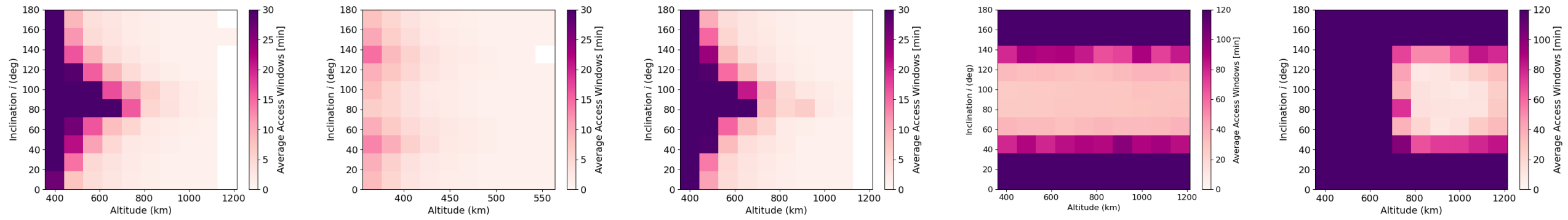

(a) OneWeb (b) Starlink (c) OneWeb+Starlink (d) Eutelsat (e) Eutelsat+OneWeb

FIGURE 5: Avg. Constellation Access Duration for (scale saturated to 30' for non-GEO systems, 120' for GEO)

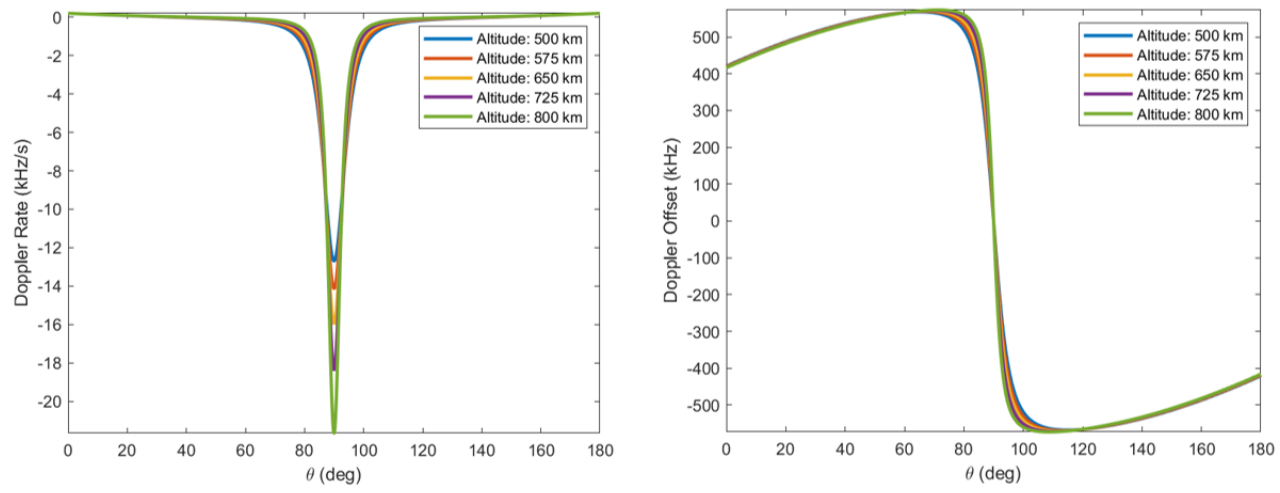

FIGURE 6: Analytical Approximation of the Space User Doppler Rate and Offset as a Function of the Elevation Angle for OneWeb Constellation

### E. FREE SPACE PATH LOSS

Figure 7 shows that average path losses range from 156 to 174 dB for OneWeb and 144 to 163 dB for Starlink, decreasing as the altitude increases because of shorter distances. On average, Starlink is 10 dB better than OneWeb for users below 500 km. However, multi-constellation users picking OneWeb and Starlink satellites at random perceive only a 6-dB improvement in the average path losses below 500 km.

Eutelsat + OneWeb path losses range from 127 dB to 206.3 dB. Selecting the closest satellite, for its reduced latency and path losses, can increase throughput and capacity, and results in average path losses of 187.41 dB, which is 7-dB better than when picking visible satellites at random. In this case, it is also possible to see how path losses vary with user orbital parameters in the LEO/GEO case: low altitude and highly inclined users have the lowest average losses because they are well covered by OneWeb and use Eutelsat only when migrating between OneWeb's orbital planes. As the user altitude increases, OneWeb's coverage decreases and the use of Eutelsat intensifies because of its higher coverage, at the cost of higher path losses.

### F. MINIMUM ELEVATION ANGLE

Coverage and availability are defined by the interplay of user and constellation orbital dynamics and their minimum elevation angles; the narrower the field-of-view of the user or the field-of-view of the constellation, that is, the higher the minimum elevation angle, the lower will be the user coverage, especially for users close to satellites moving at high relative speeds. On the other hand, lower elevation angles imply higher path losses and can be a challenge for high gain applications as a wide steering range is also necessary.

Through additional simulations (not represented), we have observed that the mean off-zenith angle is 5-to-10-degrees higher than the minimum elevation angle and that the minimum elevation angle can be increased up to 40-degrees without compromising OneWeb's coverage for most orbits. There are also small reductions in path loss ($<$ 1 dB) and Doppler ($<$ 100 kHz), but the narrower visibility cones decrease Starlink's overall coverage by 28.7%, from 37.3% to 26.6%. Consequently, Starlink's average accesses are reduced from 2.77 to 1.67 minutes. OneWeb's coverage is only reduced by 2.6%, especially at equatorial and mid inclinations, but the average access time is reduced by 35.9%, from 39.32 to 25.20 minutes.

Additionally, Eutelsat + OneWeb users experience an average uplink off-zenith angle of approximately 34.5 degrees, which corresponds to an elevation angle of 55.5 degrees. This is much greater than the 25-degree minimum elevation angle and suggests there are margins to be optimised. When one considers the 40-degree angle identified as the upper limit for the OneWeb constellation, overall coverage is reduced by 10.34%, shortening the average access by a factor of 6.8 down to 81 minutes, if not more, for users at high altitudes and inclinations because they are subject to strong geometrical constraints: too close to the OneWeb shells and too high in latitude for the GEO systems.

## VI. INTERNATIONAL SPACE STATION USE CASE

In this section, we assess the coverage that could be provided by the OneWeb and Starlink constellations to the ISS, at a 420-km altitude, 51.6 degree inclination (Figure 8). Table 5 describes the main channel parameters for an ISS-like user using Starlink, OneWeb, and their combined constellation.

| System | OneWeb | Starlink | OneWeb + Starlink |
|---|---|---|---|
| Coverage Probability [%] | 97.64 | 68.52 | 98.75 |
| Avg. Access [min] | 21.97 | 2.10 | 32.32 |
| # Visible Satellites | [0, 8] | [0, 6] | [0, 13] |
| Avg. # Visible Satellites | 2.72 | 1.25 | 3.97 |
| FSPL [dB] | [170.84, 175.88] | [154.26, 163.25] | [154.26, 175.88] |
| Avg. FSPL [dB] | 173.64 | 158.94 | 169.01 |
| Max. Doppler [kHz] | 375.65 | 506.2 | 496.92 |

TABLE 5: Key Constellation Coverage Metrics for an ISS-like User (25-degree Min. El. Angle, Random Satellite Selection)

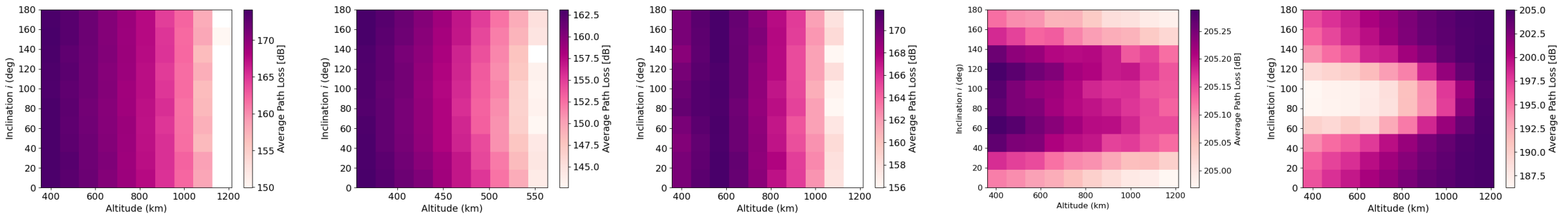

(a) OneWeb (b) Starlink (c) OneWeb+Starlink (d) Eutelsat (e) Eutelsat+OneWeb

FIGURE 7: Avg. Constellation Free Space Path Loss (Random Satellite Selection for LEO Mega-Constellations, Closest Satellite Selection for LEO+GEO.)

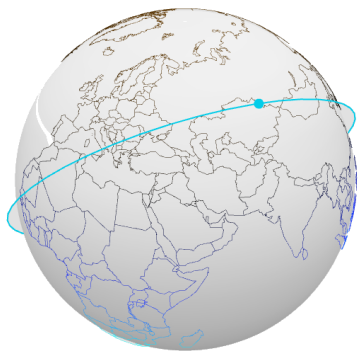

FIGURE 8: International Space Station Orbit

### A. COVERAGE

When combined, forming a 5124-satellite constellation, coverage is boosted from 97.64% for OneWeb and 68.52% for Starlink to 98.75%. Selecting the closest satellite means Starlink and OneWeb are used 68.52% and 31.48% of the simulation time, respectively, and 82% of Starlink's connectivity is through its mid-inclination shells as these satellites are closest to the ISS, also at mid-inclination (51.6-degrees). This is intuitive; Starlink satellites are much closer than OneWeb and are used whenever available. Instead, when Starlink is not visible by the user satellite because of coverage constraints, OneWeb is used. Likewise, when OneWeb is not visibile, Starlink is used. This happens much more rarely because OneWeb's coverage almost entirely encompasses Starlink's coverage.

Despite having 5124 satellites, the combined OneWeb + Starlink constellation is unable to provide continuous coverage, which shows that connectivity is inherently intermittent; even with thousands of satellites and a very high coverage probability, hours-long continuous contacts are unlikely for this user because orbital dynamics create very short, seconds long, intervals during which there are no satellites in visibility.

### B. VISIBLE SATELLITES

Notwithstanding its 4408 satellites, Starlink's coverage is lower than OneWeb's because of its greater proximity to the ISS, which exacerbates visibility constraints, reducing coverage and the number of visible satellites significantly; OneWeb has more than one satellite visible at least 50% of the simulation time, while Starlink only 33% of the time. Overall, the ISS can simultaneously see up to 6, 8, and 13 satellites for Starlink, OneWeb, and the OneWeb + Starlink, respectively, and the probability of having at least two satellites in visibility is 90% in the latter case.

### C. PASS DURATION

Contacts with Starlink satellites are very short: 87% last less than a minute, demanding much more frequent handovers than OneWeb, where only 30% of the passes last fewer than two minutes. Nevertheless, 20-minute long passes are possible when there is "coupling" between the satellites and the user. This happens for the ISS when mid-inclination shells are used; there are satellites in orbital planes that are very close to the ISS, some of which are moving towards it while others are moving away from it. Approaching satellites have a high range rate, tripling the probability of a contact shorter than one minute. However, departing satellites moving in the same direction of the ISS have low range rates, increasing pass duration. Here, it is important to highlight that this access distribution is a consequence of the satellite selection policy as picking the closest satellites can often mean selecting the satellites with the fastest relative motion, shortening pass times.

### D. NETWORK ACCESS

Considering that the constellation is accessible while there is at least one of its satellites in visibility of the user, the average constellation access lasts 21.97, 2.10, and 32.32 minutes for OneWeb, Starlink, and the combined OneWeb+Starlink constellation. Furthermore, OneWeb and Starlink orbital planes complement each other well, constellation accesses last much longer, at least 50% of them last more than 30 minutes and 15% more than one hour. It is also possible to have continuous accesses lasting more than 90 minutes when the satellites align in a way that the user is covered by the mid-inclination shell while it migrates across constellations' polar planes. This is the main contribution of the mid-inclination shells from a coverage perspective.

### E. PATH LOSS AND DOPPLER DISTRIBUTION

Figure 9 shows the path loss and Doppler offset distributions of visible satellites, that is, without applying a selection policy. If a satellite selection criterion were applied, these distributions would be re-scaled. There is a bimodal path loss distribution and, since there is no overlap, it is easy to distinguish the two constellations: 170-to-176 dB and 154-to-164 dB path losses, for OneWeb and Starlink, respectively. One can also observe that high path losses are more likely, a consequence of OneWeb's greater coverage. 70% of the

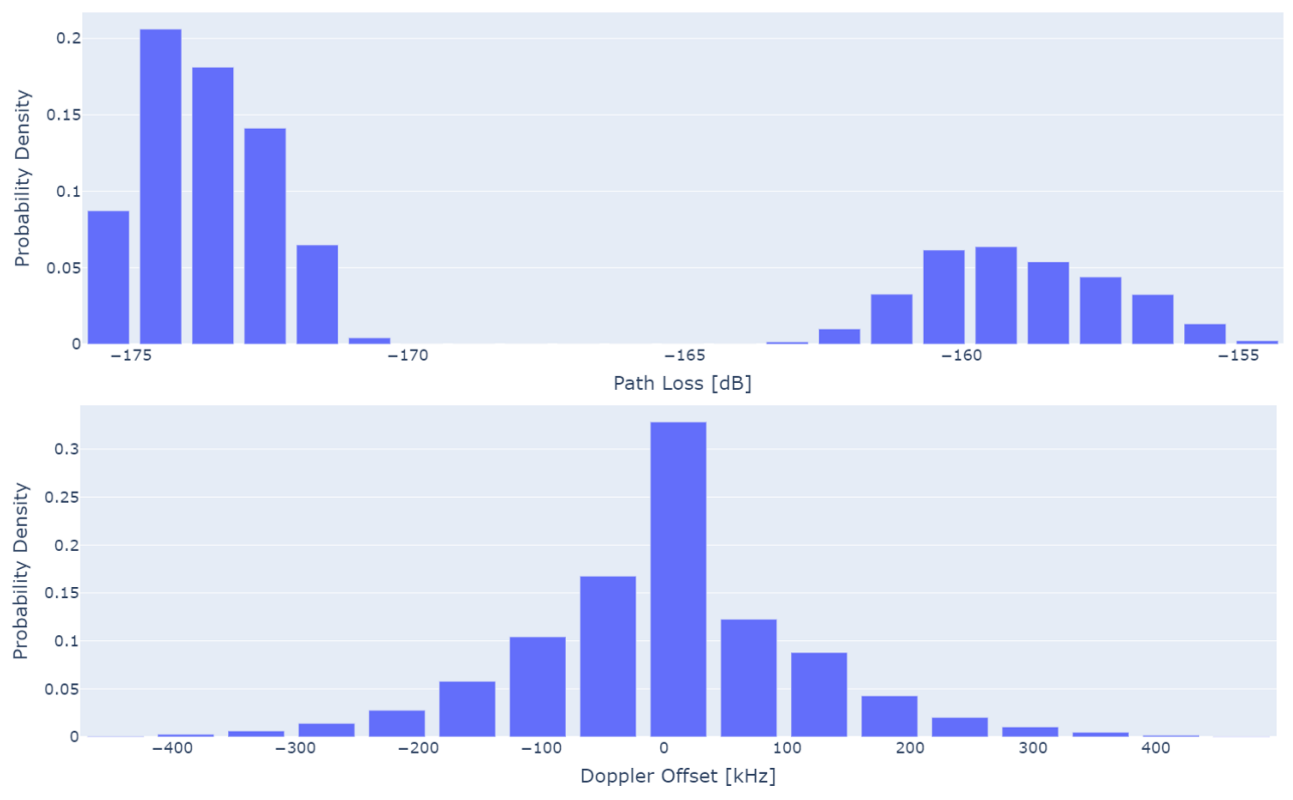

FIGURE 9: Path Loss (top) and Doppler (bottom) Distributions for an ISS-like User

Doppler offsets are lower than 100 kHz, but the maximum Doppler can reach 506 kHz when satellites are approaching from the opposite direction of the user. Curiously, while one could initially expect Starlink's lower altitudes would worsen Doppler offsets due to high range rates, the opposite happens due to the use of the mid-inclination shell where most satellites are moving at a slow rate "parallel" to the user.

### F. MINIMUM ELEVATION ANGLE

For Starlink, increasing the minimum elevation angle to 40-degrees decreases path losses by 1-dB, maximum Doppler by 100 kHz, and coverage to 49.3%, making 92.4% of accesses shorter than 2 minutes. Instead, for OneWeb, there is no considerable change in path loss or Doppler, coverage is reduced to 95.3%, and the average constellation access is reduced by 36.2%.

### G. SATELLITE SELECTION

Figure 10 shows how the number of visible satellites and the selected constellation satellites (for communication) are inherently periodic because of constellation and user orbital dynamics and the closest satellite selection criterion. Each OneWeb satellite has been labelled sequentially starting from 1. Every increase by 49 until 588 corresponds to moving from one polar plane to another. Then, from 588 to 716, since there are 8 orbital planes, an increase of 16 corresponds to a change in orbital plane in the mid-inclination shell. Starlink satellites are labelled from 589 to 5124, where a 589-to-4604 corresponds to mid-inclination satellites at 53-to-70-degrees and above 4604 to polar satellites.

For example, when the user is traversing OneWeb's polar orbital planes, it is mostly covered by one-to-two Starlink satellites. Then, as it re-enters OneWeb's coverage, the number of visible satellites increases, peaking at 13 when Starlink's and OneWeb's orbital planes align, and oscillating between two-to-three when it is covered by at least one satellite from each constellation.

### H. ADVANTAGES OF A MULTI-ORBIT SYSTEM

Using both Eutelsat and OneWeb increases coverage to 99%, boosting the average number of simultaneously visible satellites to 6.33, and average accesses to 129 minutes, allowing almost seamless connectivity.

## VII. SUN SYNCHRONOUS EARTH OBSERVATION MISSION

In this section, we assess the coverage that could be provided by the OneWeb and Starlink constellations to a 500-km altitude EO mission in SSO (Figure 11). Table 6 describes this scenario's main key performance indices.

| System | OneWeb | Starlink | OneWeb + Starlink |
|---|---|---|---|
| Coverage Probability [%] | 93.62 | 45.31 | 96.49 |
| Avg. Access [min] | 23.24 | 1.53 | 25.56 |
| # Visible Satellites | [0, 12] | [0, 6] | [0, 13] |
| Avg. # Visible Satellites | 2.59 | 0.54 | 3.14 |
| FSPL [dB] | [169.93, 174.89] | [144.82, 156.54] | [144.82, 174.91] |
| Avg. FSPL [dB] | 172.69 | 151.71 | 169.03 |
| Max. Doppler [kHz] | 456.64 | 519.9 | 519.9 |

TABLE 6: Coverage Statistics for a SSO EO User @ 500-km Altitude

### A. COVERAGE

One can see that coverage is very high when using OneWeb (93.62%) because the user's inclination is close to OneWeb's critical inclinations. Starlink, despite its polar shell, has less than half the coverage (45.31%) as its lower altitude and greater proximity to the user represent a much stronger visibility constraint. Still, Starlink can improve OneWeb's coverage by 3.06% by covering the user while it migrates between OneWeb's orbital planes.

### B. VISIBLE SATELLITES

As a consequence of their higher altitude, there are up to 12 satellites OneWeb satellites that can be seen by the user at a given time and at least two satellites on average, while only one out of Starlink's 4408 satellites is usually visible, except for short windows of time during which up to 6 satellites may be present. When combined, there are at least three visible satellites on average and up to 13 satellites in visibility. This is shown in Figure 12, which also demonstrates that the closest satellite selection criterion leads to the user periodically selecting the same satellites. It is also possible to see that the user often switches between OneWeb and Starlink satellites and uses them almost equally.

### C. PASS DURATION

Contacts between single satellites and the user are very fast due to orbital dynamics. With Starlink, 97.5% passes are shorter than one minute, while with OneWeb 90% of passes are shorter than 5 minutes. In the combined case, 70% are of passes are shorter than 1 minute and 90% last less than 5 minutes.

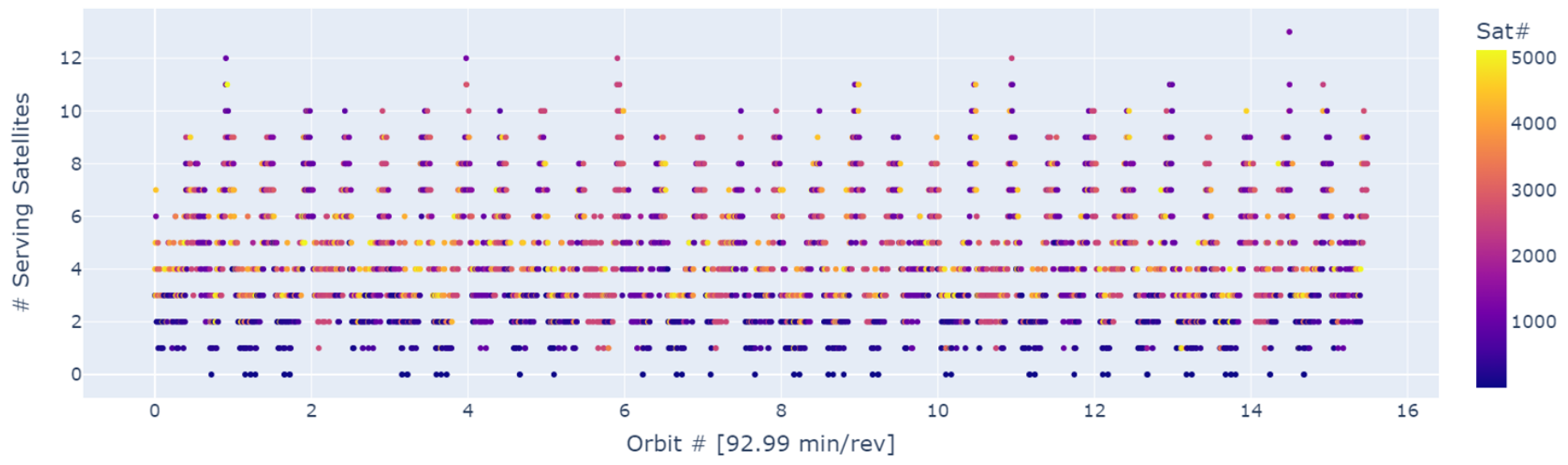

FIGURE 10: Evolution of the Number of Visible Satellites (y-axis) and Selected Constellation Satellite (label) for an ISS-like User

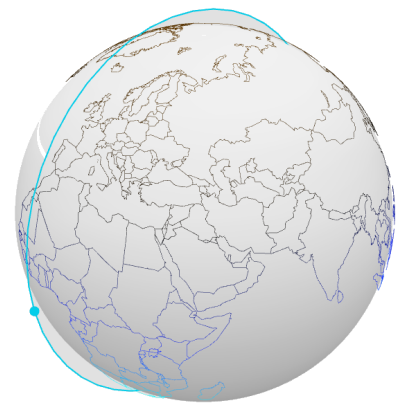

FIGURE 11: Earth Observation Mission in a 500-km Altitude Sun-Synchronous Orbit

### D. NETWORK ACCESS

Network-wise, almost all Starlink accesses are shorter than two minutes while OneWeb accesses typically last for several minutes. In the combined case, 50% of the accesses last up to 5 minutes and there a few are accesses up to a few hours long, resulting in an average access duration of 25.56 minutes.

### E. PATH LOSS AND DOPPLER

Figures 12 and 13 show that the path loss distribution is bimodal and that the user experiences 18.35-to-25.11 dB lower path losses through Starlink because of its greater proximity which may allow the use low-gain hemispherical antennas to increase coverage. By doing so, the minimum elevation angle may be as low as 5-degrees. Assessing this possibility showed that there is no improvement because at 25-degrees the elevation, the main constraint is the serving satellite beams. Nevertheless, at these short distances, it is very likely possible to exploit satellites' side-lobes (20-to-30 dB below main lobes [26]) to access space-based broadband services - an aspect to be assessed in future work. The maximum Doppler offset is on the order of 500 kHz, but most of the time it is within 200 kHz.

### F. ADVANTAGES OF A MULTI-ORBIT SYSTEM

Using both Eutelsat and OneWeb increases coverage to 99.98%, increasing the average number of simultaneously visible satellites to 8.52, and average accesses to 479 minutes. Furthermore, at least 72% of the network access of become at least 40-minutes long, nearly eliminating visibility constraints and allowing real-time applications (i.e., tasking) and data transmission.

### G. DIFFERENCES TO THE ISS USE CASE

Because the EO user is at a higher altitude (500 km) than the ISS user (430 km), Starlink coverage is reduced due to the proximity of Starlink satellites and the EO user (reduced beamwidth). Coverage falls down from 68.52% to 45.31%, penalising the average access duration, from 2.1 minutes to 1.53 minutes. On the other hand, there is almost an 8-dB reduction in the average path losses, which may prove useful in future link budgets - especially as a way to mitigate the impact of a shorter access time on capacity. Differences in OneWeb coverage of both cases are not as significant; as one expected, the EO user has a slightly lower coverage (93.62%) due to the narrower beams. Still, even if the EO user is not as well covered as the ISS user when crossing between OneWeb's orbital planes due to orbital precession, there is plenty of coverage. Moreover, despite the reduced coverage, the average access duration is one-minute longer for the EO case simply because the geometry of the problem changes; the difference in inclination between the EO user and OneWeb's polar shell is lower than in the ISS case and, therefore, the "holes" in coverage are different; there are fewer holes (longer access duration) of longer duration (coverage probability). Finally, path losses have very small differences (<1 dB) that usually fall within the design margins of such systems.

## VIII. CONCLUSIONS AND FUTURE WORK

In this work, we have presented the new concept of mega-constelation service in space. We have shown how two LEO mega-constellations in advanced deployment stages (Starlink and OneWeb) can provide space-based broadband services to LEO spacecrafts. We have described coverage, path losses, Doppler offsets, and other key parameters of the communication link as a function of space user orbital parameters. The potential advantages offered by integration with GEO satellites have been discussed. All the results have demonstrated that mega-constellations are a viable solution to limited ground station connectivity, allowing seamless connectivity to most active LEO spacecraft, as seen for two representative use cases: the International Space Station and a 500-km altitude Sun-Synchronous Earth Observation mission.

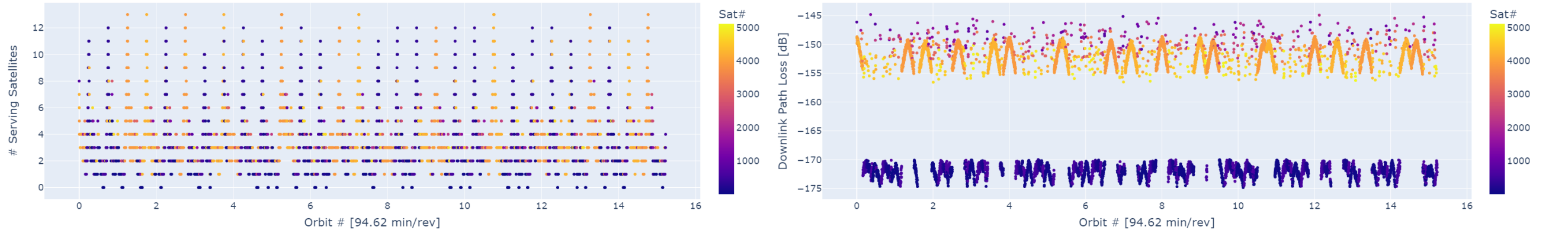

FIGURE 12: Evolution of the Number of Visible Satellites and Selected Constellation Satellite (top) and Path Losses (bottom) for a 500-km SSO User

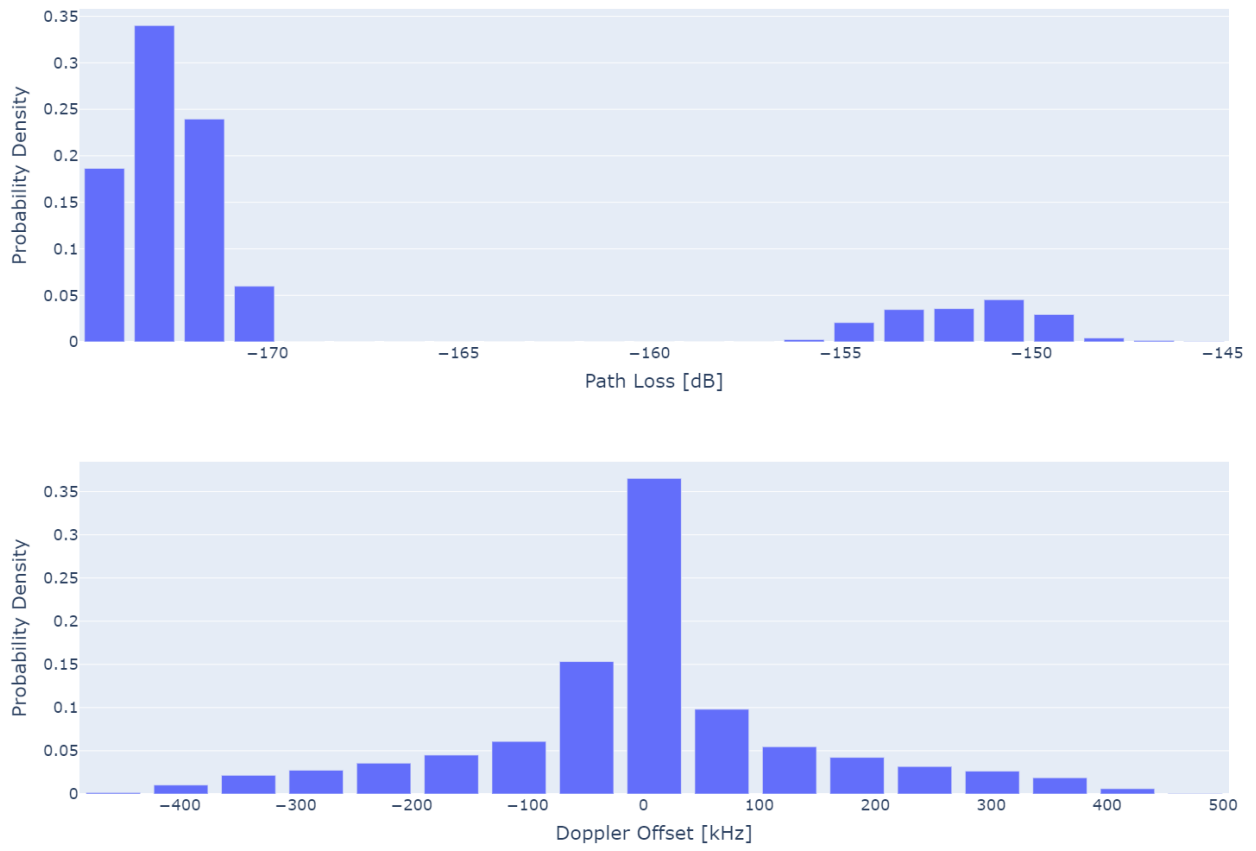

FIGURE 13: Path Loss (top) and Doppler (bottom) Distributions for a 500-km SSO User

The proposed paradigm resembles the expansion of GNSS from terrestrial services to space users but, unlike GNSS, mega-constellation services in space are much more dynamical and it is simply not sufficient to know one's position. Hence, we propose to define the mega-constellation space service volume as a function not only of a user's altitude and inclination, but also its relative position and relative dynamics (i.e., Doppler) - a definition that goes beyond the typical user position (latitude/longitude) and power constraints of GNSS. The frequent handovers and the Doppler offset requirements may represent greater operational complexity that must be accounted for when designing a space terminal. Still, because of the reduced distances between users and constellation satellites and 30-dB lower path losses, LEO mega-constellations promise reduced latency and much greater capacity than traditional GEO systems, making such a penalty worth paying.

There are several aspects of the new paradigm that can be assessed by the research community in future works. Within these, we can highlight:

1) A complete design of the LEO space terminal for deployed mega-constellations, including a performance assessment against existing services like Ground Station Networks and Data Relay Systems (on this topic we are going to present a number of results in [46]).
2) An extension of the coverage analysis to users at altitudes above 1200 km and/or eccentric orbits, and to Mid-Earth Orbit constellations.
3) An analysis on the use of side-lobes to provide service to users within 100 km of the constellation shells.
4) A detailed per-beam analysis for integrating LEO mega-constellations with the latest multi-spot GEO satellites.
5) An analysis of the impact of different satellite selection policies (i.e., at random, minimising losses, minimising Doppler, etc.), medium access and network layer aspects (i.e., orbit-aware routing and capacity allocation, congestion).

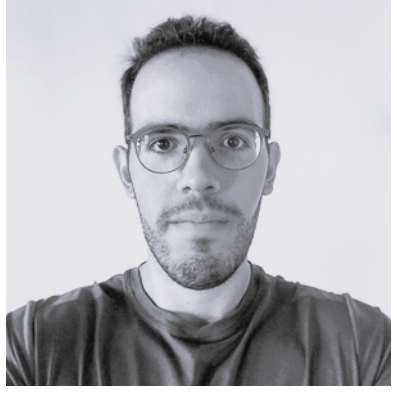

GABRIEL MAIOLINI CAPEZ is a PhD student in Electronics and Telecommunications Engineering at Politecnico di Torino, where he holds a Master's degree in Communications and Computer Networks Engineering. He has wide domain experience in satellite communication systems, space networks, and digital signal processing algorithms. His research focuses on innovative communication and ranging systems for space-to-space and space-to-ground applications, including constellation design.

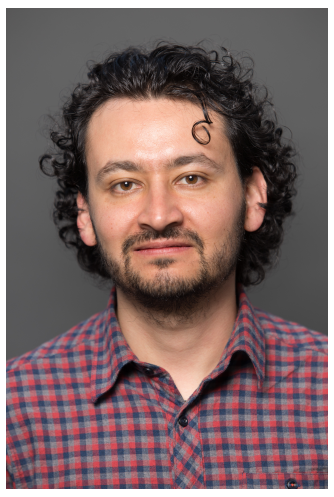

MAURICIO A. CÁCERES received a Ph.D. in Electronic Engineering in 2011 from Politecnico di Torino with a thesis on Belief Propagation for Positioning, Navigation and Timing applications with Signals of Opportunity. During his doctorate, he was a visiting student at Chalmers University, Gothenburg, and Columbia University, New York. From 2007 to 2012, he was with Istituto Superiore Mario Boella, Turin. From 2012 to 2020, he was with the German Aerospace Centre (DLR), Oberpfaffenhofen. From October 2020 he is a Senior Systems Engineer at Vyoma GmbH, Munich. His main research topics are data fusion and machine learning applied to spaceflight dynamics and space systems engineering. On these topics he has co-authored more than 20 papers. He has been the Project Manager of several research projects, and mentored PhD and Master students alike.

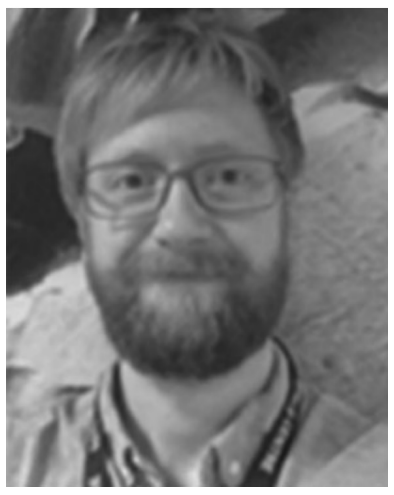

CHRIS P. BRIDGES obtained a BEng in Electronics at the University of Greenwich and was previously employed at BAE Systems in Rochester, Kent. He joined Surrey Space Centre as a PhD student in April 2006 under Dr Tanya Vladimirova and has since been successful in obtaining Post Doctorial positions in the VLSI Design & Embedded Systems and Astrodynamics Groups. He is now the On-Board Data Handling Group lead and is published in agent computing, Java processing, and multi-core system-on-a-chip technologies.

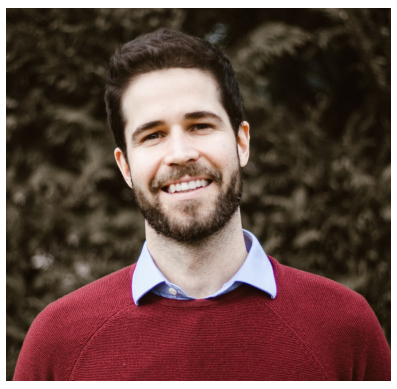

STEFAN FREY received his M.S. degree in Mechanical Engineering from ETH Zurich, Switzerland in 2014. From 2015 to 2016, he worked as a Research Assistant in the Space Debris Office of the European Space Agency. Through 2017 to 2020, he concluded his PhD studies as an Aerospace Engineer at the Politecnico di Milano, Italy, focusing on the impact that orbital fragmentation events have on active satellites in terms of collision probability. Since 2020, Dr Stefan Frey is a co-founder and managing director of Vyoma GmbH, a company focused on increasing safety in space through space-based surveillance and automation services.

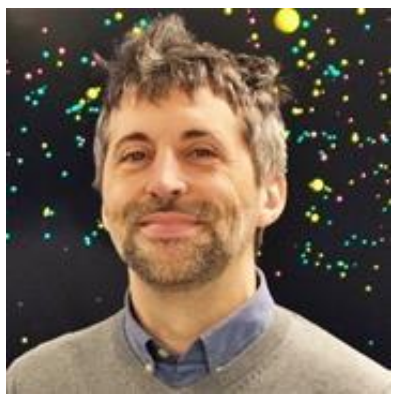

ROBERTO ARMELLIN graduated in 2003 with a first-class honors degree in Aerospace Engineering at Politecnico di Milano. In Milan, he started his research activity first as a Ph.D. student (2004-2007), and later (2007-2013) as a postdoctoral researcher developing high-order methods for problems in astrodynamics. In 2008 he co-founded the company Dinamica, for which he led (2008-2018) studies mainly with the European Space Agency. From 2014 to 2015 he was a lecturer in Astrodynamics at the University of Southampton, a position that he left when awarded the 2-year Intra-European Marie Curie Individual Fellowship Merging Lie perturbation theory and Taylor differential algebra to address space debris challenges at Universidad de la Rioja. At the end of this fellowship, he joined Surrey Space Centre at the University of Surrey as a Senior Lecturer in Spacecraft Dynamics where he also acted as Postgraduate Research Director. He spent one year as a Professor of Space Systems at ISAE-SUPAERO before joining the Te Punaha Atea Auckland Space Institute in late 2020. His main goal is now to grow New Zealand's capabilities in guidance, navigation and control, and space surveillance and tracking.

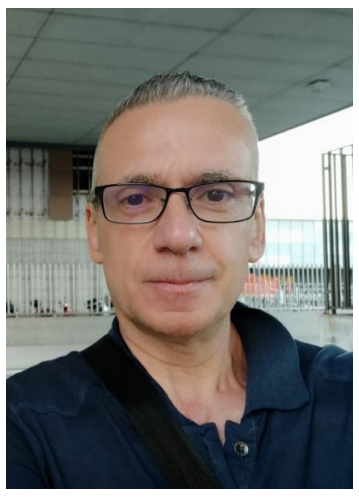

ROBERTO GARELLO received a Ph.D. in Electronic Engineering in 1994 from Politecnico di Torino with a thesis on Error Correction Coding. During his doctorate, he was a visiting student at MIT, Cambridge, and ETH, Zürich. From 1994 to 1997, he was with Marconi Communications, Genoa. From 1998 to 2001, he was an Associate Professor at the University of Ancona. From November 2001 he is an Associate Professor in the Department of Electronics and Telecommunications at Politecnico di Torino. In 2003 he was appointed Senior Member of IEEE. In 2017 he was an Adjunct Professor at California State University, Los Angeles. His main research topics are space communication systems, 5G and beyond mobile networks, and channel coding. On these topics he has co-authored more than 150 papers. He has been the Project Manager of 40 research projects, and the advisor of 12 PhD students.

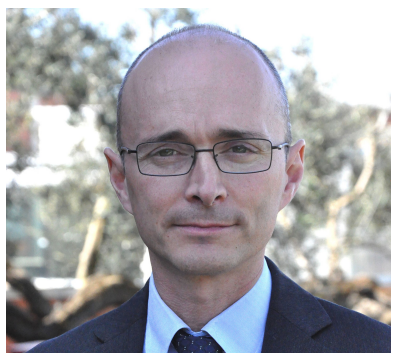

PIER BARGELLINI joined ESA in 1997 after graduating in Aerospace Engineering from the University of Pisa. For 15 years, Pier worked at the European Space Operations Centre (ESOC) in Darmstadt, Germany. During this period he was involved in the development and operations of a number of Earth observation missions. From 1997 to 2004 he was responsible for the Envisat platform operations. He was then appointed Spacecraft Operations Manager and covered this role for several EO missions, including MetOp-A's launch and early-orbit phase, ADM-Aeolus and the Sentinels. In 2012 he was nominated Head of the Copernicus Space Component (CSC) Mission Management and Ground Segment Development Division. In 2014, he became the Head of the Copernicus Space Component Mission Management and Ground Segment Development Division, within the Ground Segment and Mission Operations Department in ESA's Earth Observation Directorate. He is now the Head of the Ground Facilities Operations Division at European Space Agency.

• • •